\documentclass[draft]{aa}     
\usepackage{aabib99,graphicx}

\def\ltsim{\mathrel{\hbox{\rlap{\hbox{\lower3pt\hbox{$\sim$}}}\raise2pt\hbox{$<$}}}}
\def\gtsim{\mathrel{\hbox{\rlap{\hbox{\lower3pt\hbox{$\sim$}}}\raise2pt\hbox{$>$}}}}

\newcommand{\um}{\mbox{\,${\mu}$m}}               
\newcommand{\Msolar}{\mbox{\,M$_\odot$}}        
\newcommand{\Lsolar}{\mbox{\,L$_\odot$}}        

\newcommand{\kms}{km\,s$^{-1}$}                           
\newcommand{\Ta}{T$_{_{\rm A}}\!\!\!$*}                 
\newcommand{\co}{$^{12}$CO}                             
\newcommand{\xco}{$^{13}$CO}                            

\newcommand{\x}{\times}
\newcommand{\hyd}{H$_{2}$}                               
\newcommand{\Lco}{L$_{\rm CO}$}                        
\newcommand{\Fco}{F$_{\rm CO}$}
\newcommand{\Lbol}{L$_{\rm bol}$}                      

\def\eg{e.g.\ }
\def\ie{i.e.\ }

\begin{document}

 
\title{A Single Distance Sample of Molecular Outflows from High-Mass 
Young Stellar Objects}

\author{Naomi A.\ Ridge\inst{1}$^{,}$\inst{2} 
\and Toby J.T.\ Moore\inst{1}}         

\offprints{N.A.\ Ridge}          
                                                            
\institute{Astrophysics Research Institute, Liverpool John Moores 
University, Twelve Quays House, Egerton Wharf, Birkenhead, CH41~1LD,
United Kingdom.\\
\and
Current Address: FCRAO, 619 Lederle Graduate
Research Center, University of Massachusetts, Amherst, MA 01003, USA.
}

\mail{naomi@fcrao.umass.edu}

\date{Received  / Accepted }

\titlerunning{Single Distance Sample of Outflows}
\authorrunning{Ridge \& Moore}


\abstract{
We have made \co\ 2--1 and 1--0 maps of eleven molecular outflows
associated with intermediate to high-mass young stellar objects (YSOs)
in order to establish whether the correlations between outflow
parameters and source bolometric luminosity hold in the high-mass
regime. It is important to consider the effects of Malmquist-type
biases when looking at high-mass YSOs, as they are generally much more
distant than their low mass counterparts. We therefore chose only
objects located at $\sim$2\,kpc.  We find that the relations show much
more scatter than is seen in similar studies of low-mass YSOs. We also
find that the mass-spectrum is significantly steeper in high-mass
outflows, indicating a larger mass-fraction at lower velocities,
a low collimation factor ($\sim$ 1--2) and no Hubble-like
relationship.
}

\maketitle 

\section{Introduction}

It has been well established that outflows represent an important
evolutionary stage in the formation of low-mass stars \cite{bach96},
but systematic studies to identify bipolar outflows from massive YSOs
have only begun much more recently. There is now growing evidence that
bipolar molecular outflows are ubiquitous with YSOs of all masses.
Shepherd \& Churchwell \cite*{sc96a} searched for high velocity (HV)
CO wings toward $\sim$ 120 high-mass star forming regions, and found
that HV gas was present in $\sim$ 90\% of sources, indicating that
molecular outflows are also a common property of high-mass YSOs.  Most
luminous YSOs seem to have outflows with much wider opening angles
than their low-mass counterparts (Richer et al.\ 2000), but this may
be due to the poor resolution that the known flows have been observed
with, and/or the poor statistics available due to the small number of
flows which have been studied. It is difficult to make predictions
about the general properties of high-mass flows due to the selection
effects inherent in the small existing samples.

Shepherd \& Churchwell (1996b) determined the energetics of 5 massive
YSO outflows and showed that they fit well onto the correlation
between outflow force (\Fco, sometimes referred to as momentum flux)
and bolometric luminosity presented by \eg\ Cabrit \& Bertout (1992;
hereafter CB)\nocite{cb92}.
Henning et al.\
\cite*{henning00} studied three more massive outflows with similar
results.  However, we have shown (Moore \& Ridge, in preparation) that
these correlations, which use heterogeneous, unconstrained samples, may
be contaminated by biases due to source distance. We have therefore
undertaken a program of observations to study a larger sample of
high-mass outflows at a single distance of 2\,kpc, with consistent
sensitivity, spatial resolution and analysis. The sample is discussed
in detail in section
\ref{sectionsample}, then the observations and results are presented
in sections \ref{obs} and \ref{res}. Sections 5 and 6 contain the
discussion and conclusions respectively.

\section{The Sample}
\label{sectionsample}

We selected a base sample of 25 objects from the compilation of Wu et
al.\ \cite*{wu96}, with the selection criteria \Lbol $\ge
10^2$\Lsolar, and distance $2.0 \pm 0.3$\,kpc.  This full sample was
reduced to a final list of 11 targets by rejecting objects known to
have very weak and diffuse flows (\eg AFGL\,2591), complex and
possibly multiple flows (\eg DR\,21), and flows with little obvious
bipolarity (\ie flows which may be pole-on, for which it is difficult
to define the flow extent).  Table \ref{sdsample} lists the selected
outflows. All of these flows have been previously observed with low
resolution and with different sensitivities, but few have been
previously mapped in any detail.
\begin{table*}
\begin{center}
\begin{tabular}{ccccccc}
\hline
Source      & R.A.(1950)& Dec.(1950)& \Lbol     & D      &V$_{\rm LSR}$& Refs$^b$\\
            & hh:mm:ss  & dd:mm:ss  & \Lsolar   & kpc    & \kms & \\
\hline
NGC\,6334B & 17:17:18 & --35:48:00 & $4.00\x10^5$ & 1.7 &0& 1\\
NGC\,6334I & 17:17:34 & --35:43:47 & $8.00\x10^4$ & 1.74 &--7.5& 1\\
GGD\,27 & 18:16:13 & --20:48:43 & $2.0\x10^4$ & 1.7 &+12& 1\\ 
S88\,B &19:44:40 & +25:05:30 & $1.80\x10^5$ & 2.0 &+21& 1\\ 
IRAS\,19550 &19:54:59 & +32:48:29 & $1.50\x10^2$ & 2.0 &+12& 1\\ 
IRAS\,20188 &20:18:51 & +39:28:18 & $1.26\x10^4$ & 2.0 &+2& 1\\ 
W75\,N & 20:36:51 &+42:27:20 & $1.40\x10^5$ & 2.0 &+9& 1,2\\ 
W3\,IRS5 & 02:21:53 &+61:52:21 & $1.10\x10^6$ & 2.3$^a$ &--40& 1\\
AFGL\,437   & 03:03:32 & +58:19:37 & $2.40\x10^4$  & 2.0   &--39& 1,3\\
AFGL\,5142  & 05:27:28 & +33:45:37 & $3.80\x10^3$  & 1.8   &--4& 1,4\\
AFGL\,5157  & 05:34:33 & +31:57:40 & $5.50\x10^3$  & 1.8   &--18& 1\\
\hline
\end{tabular}\\
\end{center}
\footnotesize{$^a$Imai et al.\ \cite*{imai00} find a distance to W3\,IRS5 of 
1.8$\pm$0.14\,kpc based on maser observations.  $^b$Refs: 1.~Wu
et al.\ \cite*{wu96}, 2.~Moore \cite*{moore89}, 3.~Weintraub \& Kastner\cite*{wk96}, 4.~Hunter et al.\ \cite*{hunter95}.}
\begin{center}
\caption[Single distance sample of outflow sources]
{Single distance sample of outflow sources. Columns 2 \& 3 give the
positions of the sources. All offsets quoted throughout this paper are
given relative to these positions. V$_{\rm LSR}$ is the velocity of
the ambient cloud material relative to the local standard of
rest. \Lbol\ is the bolometric luminosity of the source believed to be
the driving source for the outflow. Luminosities given by Wu et al.\ (1996)
were checked by referring to the original paper, and found to be unreliable
in several cases.
}
\end{center}
\label{sdsample}
\end{table*}
The sources are selected from an existing, heterogeneous catalogue of
outflow sources, and therefore may not be truly representative of the
population properties. However, by selecting objects with a range of
luminosities, all at the same distance, biases should be reduced to
minimum. The bolometric luminosities listed in Table \ref{sdsample}
are generally based on IRAS fluxes and thought to be reliable. 

\section{Observations and Data Analysis}
\label{obs}
Observations were carried out between 1998 August and 1999 June at the
15m James Clerk Maxwell Telescope (JCMT) on Mauna Kea, Hawaii and at
the NRAO 12m Radio Telescope (NRAO) at Kitt Peak, Arizona.

\subsection{JCMT Observations}
We obtained \co\ J=2$\rightarrow$1 ``flexibly-scheduled'' observations
between 1998 August and 1999 May and 4 scheduled nights of
observations in 1999 July. At the observing frequency of 230\,GHz the
half-power beam width of the telescope is 21$''$. We used the DAS
spectrometer with 1657 channels, providing an effective velocity
resolution of 0.4\,\kms. The system temperatures during the
observations varied between 300\,K and 500\,K depending on the
weather. Pointing and focus checks were carried out regularly on
planets or bright galactic sources. Pointing was found to vary by less
than 5$''$ during the runs.

The observations were carried out in raster-mapping mode with a grid
spacing of 10$''$ producing fully sampled maps.  In this mode, the
telescope observes one ``off'' position per row of on-source
observations.  Maps were repeated 4 times to obtain the required rms
sensitivity of 0.2\,K per 1\,\kms\ resolution element. The reference
positions were checked for the presence of CO emission by single
position-switched observations.

The data were calibrated using the standard chopper-wheel technique,
so the intensity scale for all JCMT spectral data presented here is
expressed in units of \Ta\ \cite{ku81} which is the source
antenna temperature corrected for atmospheric and rearward spillover
losses.  This is converted to radiation temperature, by correcting for
the forward spillover efficiency $\eta_{fss}$.  Single
position-switched observations of a standard source (W3) were also
obtained so that the data could be compared with the NRAO
observations.

\subsection{NRAO Observations}
We obtained \xco\ J=2--1 and \co\ J=1--0 observations at the NRAO 12m
telescope during scheduled runs in 1998 November, 1999 March and 1999
April. These data were necessary in order to investigate
variations in optical depth and excitation which can significantly
affect the masses derived via the LTE approximation.

The telescope has a half-power beam width of 27$''$ at 230\,GHz
and 55$''$ at 115\,GHz.  At 230\,GHz the dual-channel single-beam SIS
receiver was used with two filter bank spectrometers in parallel, each
with 128 channels. The filter banks were used at 500\,kHz and 1\,MHz
bandwidths providing a spectral resolution of 0.7\,\kms\ and
1.3\,\kms\ respectively at 230\,GHz.  System temperatures varied
during the runs, and channel 2 of the receiver achieved consistently
worse performance during the observations.

The observations were made in On-the-Fly mapping mode.  As at
the JCMT, the chopper wheel calibration method was used, yielding
intensities on the \Ta\ scale.

The mapped fields were the same size as the JCMT fields, plus a 30$''$
``ramp-up'' distance added to the end of each row. A row spacing of
8$''$ and row scanning rate 44$''$s$^{-1}$ were used, resulting in
fully-sampled maps.  Maps were repeated, scanning alternately in the
RA and Dec directions to remove artifacts in the scanning direction,
and later combined and gridded together to improve signal-to-noise in
the final map.

\subsection{Analysis}

A local thermal equilibrium (LTE) approximation was used to convert the
\co\ intensities into column densities, using the \xco\ 2--1 data to
correct for optical depth variations with velocity and position,
assuming a solar-system isotopic abundance ratio of 89. Smaller
values of this ratio have been observed by \eg Langer \& Penzias
\cite*{lp90} but this will affect the derived masses by less than a
factor of two \cite{thesis}.
\co\ 1--0 data
was additionally used to calculate excitation temperatures to improve
the analysis.
\co\ column densities were then converted to masses, using a standard
$\left[\frac{X_{CO}}{X_{H_2}}\right]$ abundance Galactic ratio.

Flow dynamics were derived following the method of Cabrit \& Bertout
(1992).  A full description of the data reduction and analysis
procedure is given in Ridge \cite*{thesis}.

\section{Observational Results}
\label{res}
\subsection{Morphology}
\label{contour}

\begin{figure*}
\vspace*{19cm}
\includegraphics{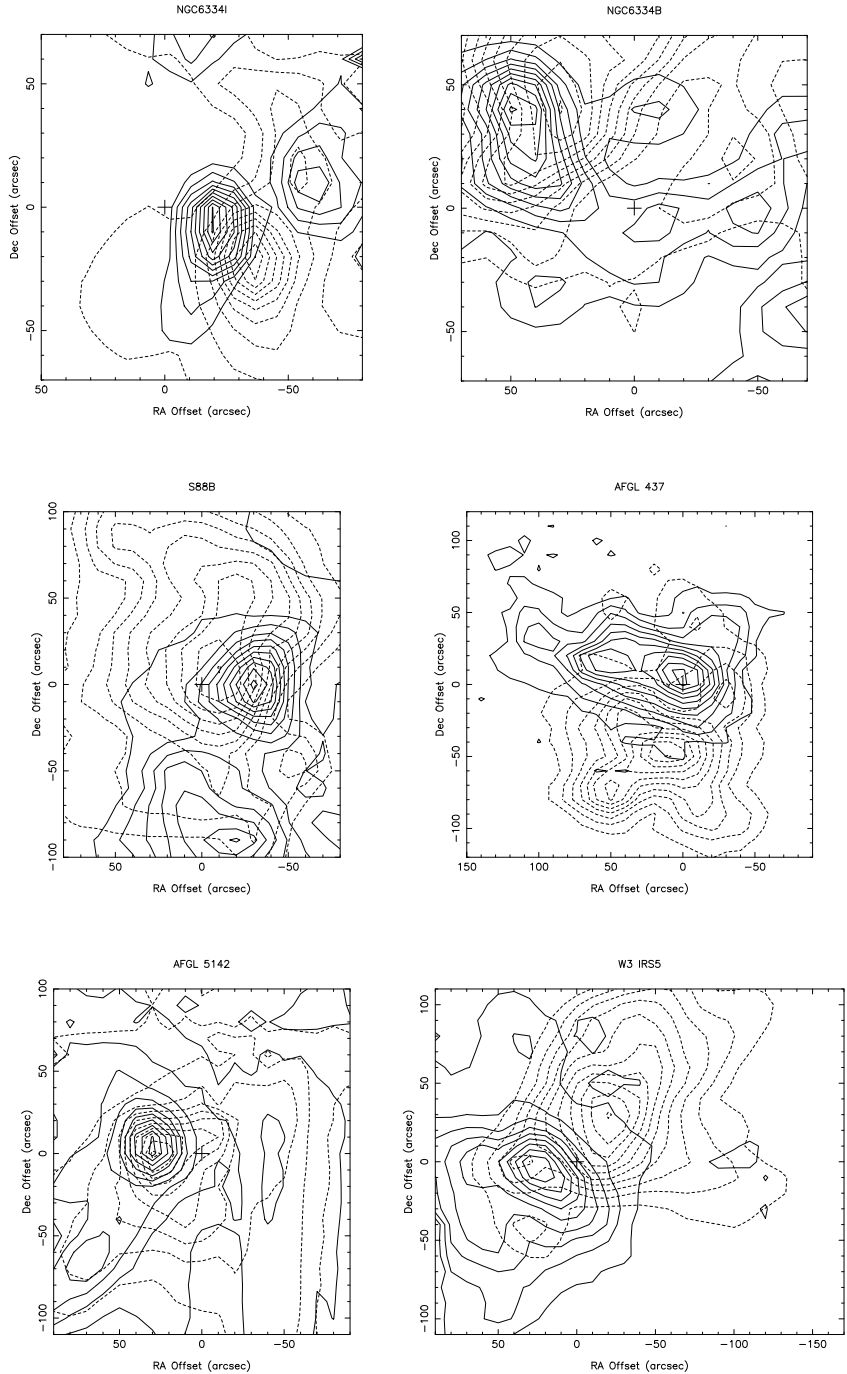}
\caption{Maps of the integrated \co\ J=2--1 emission from 6 sources 
made at the JCMT with 21$''$ resolution. Contours are at 10\%
intervals in all cases.  Blue-shifted material is indicated by dashed
contours and red-shifted material by solid contours. Crosses indicate
the position of the infrared source associated with each of the flows.
Velocity integration ranges are given in Table \ref{vels}.}
\label{contour1}
\end{figure*}

Maps of the \co\ J=2--1 integrated intensity from JCMT are shown in
Figs. \ref{contour1} and \ref{contour2}.  Blue-shifted emission is
shown as dashed lines, and red-shifted emission is shown as solid
lines.  Contour levels are as given in the captions, with the
peak-flux (integrated \co\ flux at the position of maximum integrated
intensity) summarised in Table \ref{fluxes}.  The integration limits
for the red and blue shifted gas are summarised in Table \ref{vels}.
In general the flows are not well-collimated, but bipolar structure is
evident in most cases. We discuss each object individually below.
\begin{table}
\begin{center}
\begin{tabular}{ccc}
Source & F$_{\rm peak}$(red) & F$_{\rm peak}$(blue)\\
       & K\,km\,s$^{-1}$ & K\,km\,s$^{-1}$\\
\hline
NGC\,6334I & 290 & 640\\
NGC\,6334B & 214 & 252\\
S88\,B & 130 & 240\\
AFGL\,437 & 130 & 100\\
AFGL\,5142 & 195 & 163\\
AFGL\,5157 & 168 & 81\\
W3\,IRS5 & 680 & 656\\
GGD\,27 & 100 & 190\\
IRAS\,19550 & 40 & 36\\
IRAS\,20188 & 222 & 245\\
W75\,N & 183 & 60\\
\hline
\end{tabular}
\end{center}
\caption{\co\ flux at the positions of maximum integrated
intensity of the red- and blue-shifted emission.}
\label{fluxes}
\end{table}
\begin{table}
\begin{center}
\begin{tabular}{ccc}
Source & $\Delta V_{\rm red}$& $\Delta V_{\rm blue}$\\
& \kms & \kms\\
\hline
NGC\,6334I & 2.5$\rightarrow$ 40.0& -40.0$\rightarrow$ -10.0\\
NGC\,6334B &3.0$\rightarrow$ 6.0 & -18.0$\rightarrow$ -6.0\\
S88\,B & 24.5$\rightarrow$ 31.0&7.5$\rightarrow$ 20.0 \\
AFGL\,437 & -37.0$\rightarrow$ -25.0&-48.0 -43.0 \\
AFGL\,5142 &-3.0$\rightarrow$ 6.0 &-17$\rightarrow$ -5 \\
AFGL\,5157 &-16.0$\rightarrow$ -3.0 &-30.0$\rightarrow$ -21.5\\
W3\,IRS5 &-39.0$\rightarrow$ -14.0 &-61.0$\rightarrow$ -43.0 \\
GGD\,27 &14.5$\rightarrow$ 20.0 &1.5$\rightarrow$ 10.5 \\
IRAS\,19550 &14.0$\rightarrow$ 21.0 &5.5$\rightarrow$ 11.5 \\
IRAS\,20188 &4.0$\rightarrow$ 16.0 &-20.0$\rightarrow$ -2.0 \\
W75\,N &15.0$\rightarrow$ 32.5  &-30.0$\rightarrow$ -18.0 \\
\hline
\end{tabular}
\end{center}
\caption{The ranges of red and blue shifted emission over which
the integration was carried out. Velocities are LSR.}
\label{vels}
\end{table}

\subsubsection{NGC\,6334I}
NGC\,6334 is a large complex of H\,{\sc ii} regions and an associated
molecular cloud in the southern Galactic plane, containing numerous
sites of recent or ongoing star formation.  NGC\,6334I is the
brightest far-infrared source in the northern part of NGC\,6334
\cite{lough} and is the dominant source at millimetre and 
sub-millimetre wavelengths \cite{sandell}. The high-velocity outflow
was first detected by Bachiller \& Cernicharo \cite*{bc90}. NGC\,6334I
is also a rich source of molecules \cite{bc90} and has water, methanol
and OH masers associated with it. Three ammonia masers have also been
detected near the ends of the outflows \cite{kj95}.

The integrated contour map of the NGC\,6334I outflow is shown in
Fig.\ \ref{contour1}. There are two distinct flows, indicated by two
clearly separated lobes present in the red-shifted emission (solid
contours). It is less clear in the blue-shifted emission, but there
seems to be an extension to the north-west which could be a second
flow lobe.  The second, less-dominant flow seems to be centred to the
west of the stronger flow, and probably originates from a second
object. This is confirmed by McCutcheon et al.\ \cite*{mccutcheon} who
detect this feature in both \co\ 3--2 and in CS, and assign it the
name I(NW).  I(N) is a third object in this region, located $\sim 100''$
north of NGC\,6334I. The bottom edge of the outflow from this source can
be seen on the northern edge of the frame.
Outflow dynamics were calculated from the
dominant outflow, closest to the IRAS source.
Extremely high velocity emission is seen in the spectra of this source
and there is a sharp absorption feature at V$_{\rm LSR}$
+6.5\,\kms. This is a real line in the same sideband, not due to
emission in the off position and is also seen in J=3--2 emission by
McCutcheon et al.\
\cite*{mccutcheon} widespread over the northern part of the NGC\,6334
complex, suggesting that a foreground cloud at a lower temperature may
be obscuring the radiation from NGC\,6334I.

\subsubsection{NGC\,6334B}
The second source in the NGC\,6334 complex, NGC\,6334B was observed by
Phillips \& Mampaso \cite*{pm91}. Their observations centred on the CO
peak position of Dickel et al.\
\cite*{ddw77}, and they did not find a bipolar structure in the 
high-velocity gas. However, there is an extremely luminous FIR source
(L$\sim 4\times10^5$\Lsolar; Loughran et al.\ 1986\nocite{lough}) and
H\,{\sc ii} region 40$''$ to the north of this position, and our maps
were therefore offset north 40$''$ from the CO peak position.

The integrated contour map of the NGC\,6334B outflow is shown in
Fig.\ \ref{contour1}. This flow shows two lobes well offset to the
north-east from the centre of the map. The peak emission of the two
lobes is separated by one beam width. The blue lobe may stretch beyond
the map boundary to the north.  The flow lobes are not well
collimated. Both the \co\ and \xco\ spectra show peaks at 0 and
-5\,\kms, indicating two separate cloud components, and the spectrum
from the red peak position shows a possible third peak at
3\,\kms. This is clearly a complex region.

\subsubsection{S88\,B}
{S88\,B} is a compact H\,{\sc ii} region located within the S\,88
cloud. Single-dish molecular line observations show that the S\,88B
complex is embedded within a molecular cloud core of relatively low
mean density, $n(H_2)\sim 6 \times 10^3$\,cm$^{-3}$, out of which the
exciting stars of the H\,{\sc ii} regions presumably formed (\eg
Phillips \& Mampaso 1991 \nocite{pm91} and refs. therein). The outflow
was mapped by Phillips \& Mampaso \cite*{pm91} with 30$''$ resolution.

The integrated contour map of the outflow from S88\,B is shown in Fig.\
\ref{contour1}. Two lobes are clearly visible, with the red-shifted lobe
(solid contours) appearing more collimated than the blue-shifted lobe
(dashed contours). Two distinct peaks are present in the red-shifted
emission, but the blue-shifted emission gives no indication of a
second bipolar outflow to the south.

\subsubsection{AFGL\,437}
AFGL\,437 is a compact cluster \cite{klein77} containing at least 3
highly polarised YSOs \cite{dyck79}. Asymmetric profiles were detected
near this source by Schneps et al.\ \cite*{schneps}, but were not
initially interpreted as a bipolar outflow. Arquilla \&
Goldsmith\cite*{ag84} made observations at 50$''$ resolution and a
higher sensitivity and interpreted the profiles as produced by a
large-scale bipolar outflow, with the red and blue lobes located at
the north and south of the central cluster, and their maxima
separated by $\sim 1.2'$. G{\'o}mez et al.\ \cite*{gomez92} mapped the
region at higher resolution and found a compact outflow with a very
low degree of collimation.

The integrated contour map of AFGL\,437 is shown in Fig.\
\ref{contour1}.  This is a poorly collimated compact flow. However
bipolar structure is still evident, with the blue lobe stretching
south and the red lobe extending to the north. The north lobe
particularly appears elongated perpendicular to the outflow axis.  The
AFGL\,437 infra-red cluster, indicated in the diagram by the cross is
not at the centre of the outflow.  We suggest that the contour pattern
here is consistent with the base of a very wide parabolic outflow in a
clumpy medium.
Weintraub \& Kastner \cite*{wk96} use infrared
polarisation data to explain the unusual morphology of this CO outflow
by the presence of an obstruction to the north of AFGL\,437 which will
defocus and deflect the flow. It may also be due to the position of
the cluster near to the edge of the molecular cloud \cite{gomez92}.

\subsubsection{AFGL\,5142}
The star-forming region AFGL\,5142 contains an IRAS source with a
bolometric luminosity 3.8$\times 10^3$\Lsolar\ \cite{carp90}. The CO
outflow was found by Snell et al.\ \cite*{snell88} with moderate resolution
observations. Its radio continuum flux density is consistent with a
B2\,ZAMS star, assuming optically thin free-free emission
\cite{torrelles92a}.

There is a cluster of infrared sources in the centre of the region
with a radius of 0.3\,pc (30$''$ at 2 kpc; Hunter et al.\
1995\nocite{hunter95}), many of the sources showing a strong infrared
excess. Two cluster members, IRS\,1 which has the strongest IR excess
and IRS\,2, the brightest member at K-band (2.7 magnitudes brighter
than IRS\,1) have received the most attention in previous
studies. IRS\,2 is at the centre of the large scale outflow
\cite{hunter95} and also coincides with the IRAS point source position
(marked with a cross in the figure). Hunter et al.\ suggest that
IRS\,2 is in an advanced stage of protostellar evolution, in a region
where the molecular cloud has been dispersed. IRS\,1 is thought to be
in a much earlier evolutionary phase, coinciding with an unresolved
thermal continuum source, a cluster of H$_2$O masers and a small dense
molecular cloud core.

The integrated contour map of AFGL\,5142 is shown in Fig.\
\ref{contour1}.  This is another relatively compact flow, with the
blue and red peaks superimposed suggesting a flow at a small angle to
the line of sight. This flow is poorly collimated, and Hunter et al.\
\cite*{hunter95} suggest it may be multi-polar, with two flows
orientated almost perpendicular to one another. Our observations
do not support this suggestion.
\subsubsection{W3\,IRS5}
The source IRS\,5 in the W3 giant molecular cloud was discovered by
Wynn-Williams et al.\ \cite*{ww72}. It has a very high luminosity
($\sim 10^6$\Lsolar), extremely steep spectrum and a very deep
silicate absorption feature at 9.8\,\um\ \cite{willner77}. The outflow
was first identified by
Bally \& Lada \cite*{bl83}.

The integrated contour map of W3\,IRS5 is shown in Fig.\
\ref{contour1}.  Although relatively poorly collimated, bipolar structure
is clearly evident in this outflow. Emission is present out to
$\gtsim$ 20\,\kms\ in this flow.  This source shows strong
self-absorption, especially noticeable in the spectrum from the blue
lobe, where the line centre, indicated by the peak of the \xco\
spectrum, is almost totally self-absorbed.
\begin{figure*}
\vspace*{19cm}
\includegraphics{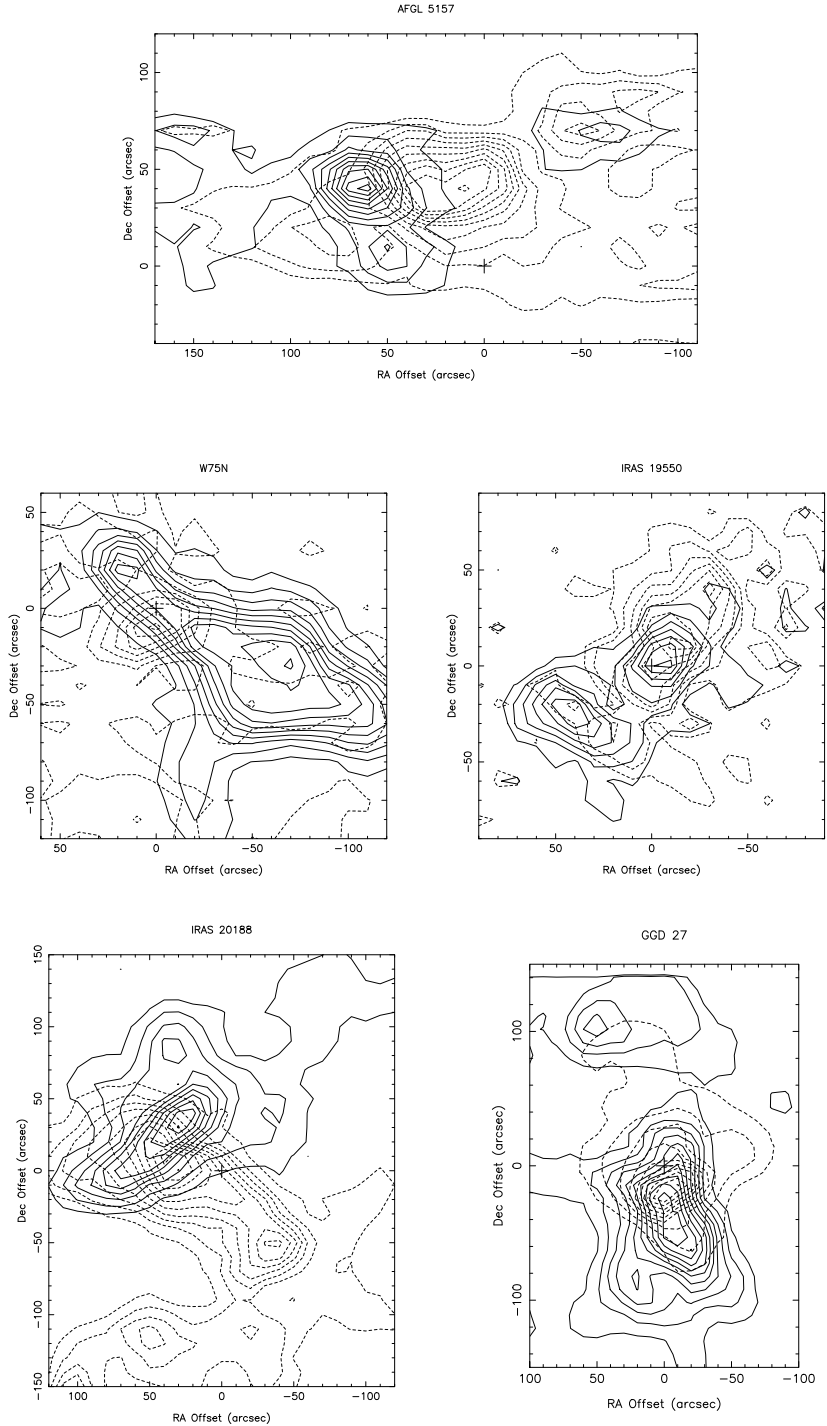}
\caption{Maps of the integrated \co\ J=2--1 emission from 5 sources
made at the JCMT with 21$''$ resolution.
Contours are at 10\% intervals in all cases.
Blue-shifted material is indicated by dashed contours and red-shifted
material by solid contours. Crosses indicate the position of the
infrared source associated with each of the flows. Integration limits 
for the red and blue wings are given in Table \ref{vels}.}
\label{contour2}
\end{figure*}
\subsubsection{AFGL\,5157}
AFGL\,5157, situated in the Perseus arm \cite{roberts72}, is an
extensively studied star-forming region. Two optical reflection
nebulae, NGC\,1985 and GM\,39, have been observed in AFGL\,5157 (Higgs
1971\nocite{higgs71}, Torrelles et al.\ 1992a\nocite{torrelles92a}).
Near-infrared observations by Torrelles et al.\ \cite*{torrelles92b}
presented two infrared sources, IRS 1 and 2. IRS\,1 appears to be
associated with the infrared source IRAS 05345+3157, which has a
far-infrared luminosity of 5.5 $\times 10^3$
\Lsolar\ \cite{snell88}. IRS\,2 is located close to the reflection
nebula GM\,39. In the K$'$ imaging survey by Hodapp \cite*{hod94}, a
cluster of stars can be seen in the IRS\,1 position associated with
dispersed nebulosity.

The IRAS source 05345+3157 is not thought to be the powering source of
the outflow in AFGL\,5157, as it is located $\sim1'$ southwest of the
outflow and molecular core centre \cite{verdes89}. Torrelles et al.\
\cite*{torrelles92b} suggest that the outflow is in fact powered by a
deeply embedded source at the centre of the ammonia core and
associated with the strong H$_2$O maser.

The integrated contour map of AFGL\,5157 is shown in Fig.\
\ref{contour2}.  This flow shows a high degree of collimation and
bipolarity with weak extensions stretching for an arcminute to the
east and west. The majority of the emission is however concentrated
close to central source. There may be a weak second flow located to the
north-west.
\subsubsection{W75\,N}
W75\,N is a well studied active star-forming region containing compact
H\,{\sc ii} regions, OH and water masers, 2\um\ emission and about 40
embedded sources, as well as the well-known CO outflow (Moore et al.\
1991a, \nocite{moore91a} Moore et al.\ 1991b\nocite{moore91b}).  It is
thought to be forming several B-type stars \cite{hunter94}.  The
source associated with IRS\,1 and H\,{\sc ii} region 'B', with a
luminosity of $\sim1.4\times 10^5$\Lsolar\ \cite{moore91a} is thought
to be driving the outflow
\cite{moore91b}. H\,{\sc ii}(B) has since been resolved into three
subcomponents by Hunter et al.\ \cite*{hunter94}.

The integrated contour map of W75\,N is shown in Fig.\ \ref{contour2}.
There is a well collimated red-shifted lobe which stretches south west
from the source. It appears to start very narrow, then open up as it
gets further from the source.

It is more difficult to determine the morphology of the blue-shifted
lobe, as the spectra are contaminated by emission from the nearby
cloud containing the object DR\,21 which has an LSR velocity of
$-$3\,\kms\ and overlaps the W75 cloud over most its extent (Dickel,
Dickel \& Wilson 1978). The \xco\ spectra show the two separate
objects clearly, but it is impossible to determine how much of the
\co\ emission is from the wings of W75\,N and how much from DR\,21.
The peak at an offset of (0,$-$15) arcseconds is strongest at
$\sim$+3\,\kms\ and is therefore likely to be due to the outflow.
However, this peak is compact, showing virtually no extension along
the axis of the red outflow lobe. Davis et al.\ \cite*{davis98b}
suggested that the small extent of the blue lobe is probably due to
environmental asymmetry, with little molecular gas to the north-east
of the source, and supported by the fact that the asymmetry of the
flow is not reflected in \hyd\ observations.  Due to the unusual
nature of the blue-lobe of this flow, all calculations of flow
properties were made by scaling from the red-lobe values.
\subsubsection{IRAS\,19550}
IRAS\,19550 was identified as a protostellar object from its outflow
by Koo et al.\ \cite*{koo94}. This object appears to be forming in
isolation at the tip of a large filamentary cloud. IRAS\,19550 shows
K-band emission extended along the east-west direction, and I and R
band images show that the source is made up of two weak peaks shifted
symmetrically from the K-band peak, probably due to scattered stellar
light \cite{koo94}.

The integrated contour map of IRAS\,19550 is shown in Fig.\
\ref{contour2}. This flow is the most compact of those we studied. 
There is also some indication of a second flow located to the
south-west of the source and at right angles to the known flow,
particularly evident in the red (solid) contours.  The source has
very weak \co\ emission.
\subsubsection{IRAS\,20188}
IRAS\,20188, a compact molecular cloud identified in a survey for
molecular line emission from IRAS sources by Richards et al.\
\cite*{richards87}, is located in the Cygnus region and is not
obviously associated with any known stellar cluster or H\,{\sc ii}
region. The outflow was first identified by Little et al.\
\cite*{little88}.

The integrated contour map of IRAS\,20188 is shown in Fig.\
\ref{contour2}. The morphology of this flow is difficult to decipher.
There appear to be two well-collimated lobes at almost 90 degrees to
each other, both containing at least two distinct peaks, which may
suggest a multiple flow or alternatively may be a result of multiple
outflow episodes. One explanation of this flow is that it is the base of 
a very wide parabolic flow with an asymmetric distribution of material.
\subsubsection{GGD\,27}
GGD\,27 was originally catalogued as an optical candidate Herbig-Haro
object \cite{gyul78}, but has now been determined to be the centre of
a highly active star-forming region. It contains numerous
near-infrared point sources and extensive reflection nebulosity
\cite{yama87}. The deeply embedded object GGD\,27-ILL is thought 
to be the driving source of the outflow \cite{aspin94}, and the nearby
source IRS\,2 is a knot of dust heated by the the hot circumstellar
dust around GGD\,27-ILL.

The integrated contour map of GGD\,27 is shown in Fig.\
\ref{contour2}.  There are two lobes well separated in velocity (blue
lobe integrated between 1.5\,\kms\ and 10.5\,\kms, red lobe integrated
between 14.5\,\kms\ and 20.0\,\kms), although they are superimposed
in space, with the blue lobe stretching to the north, while the red
lobe stretches southwards. The red lobe (solid contours) appears
slightly more collimated than the blue lobe (dashed contours).

\subsection{Outflow Properties}

\subsubsection{Flow Sizes and Dynamical Timescales}

\begin{figure*}
\vspace*{7cm}
\includegraphics{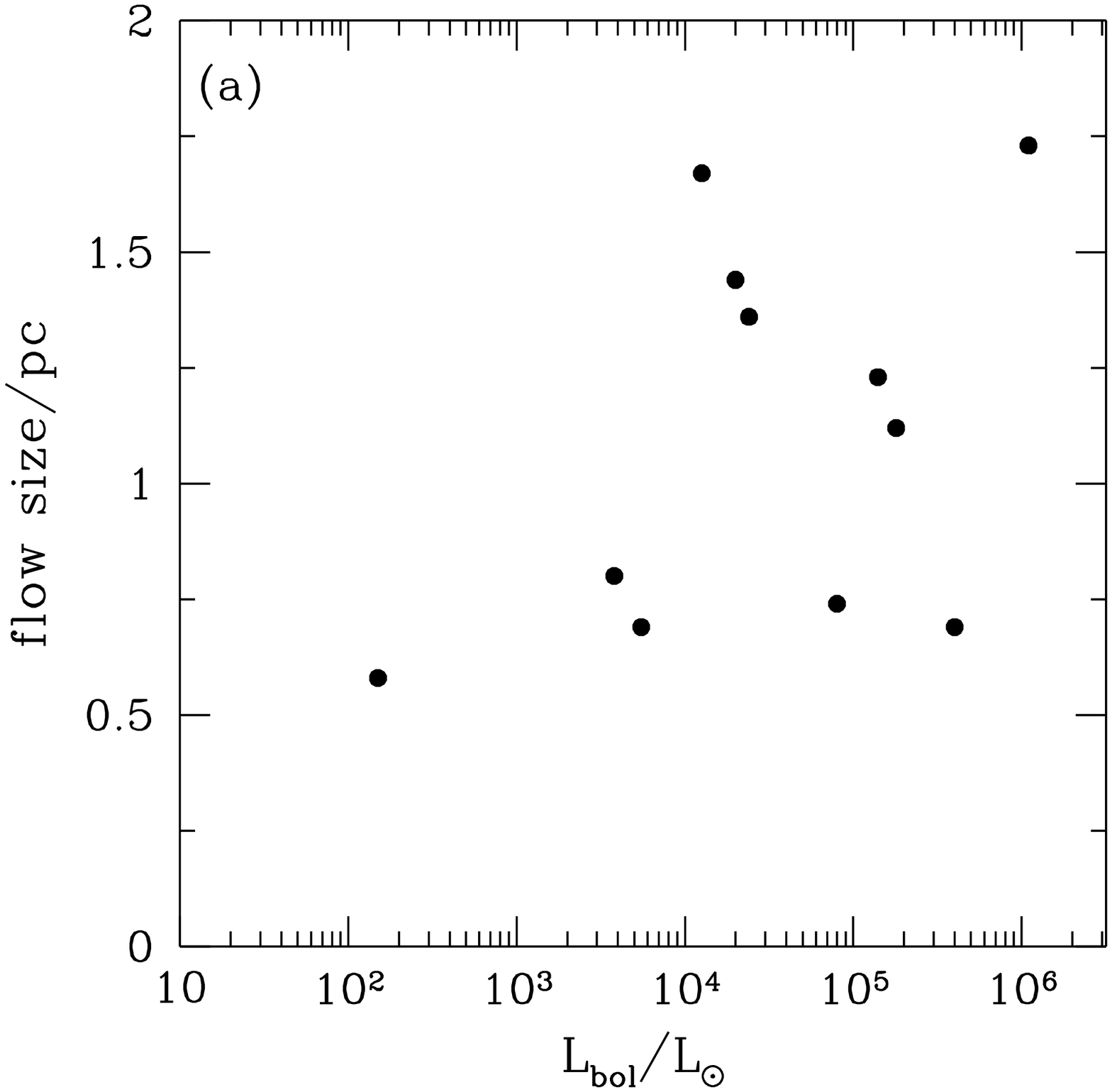}
\includegraphics{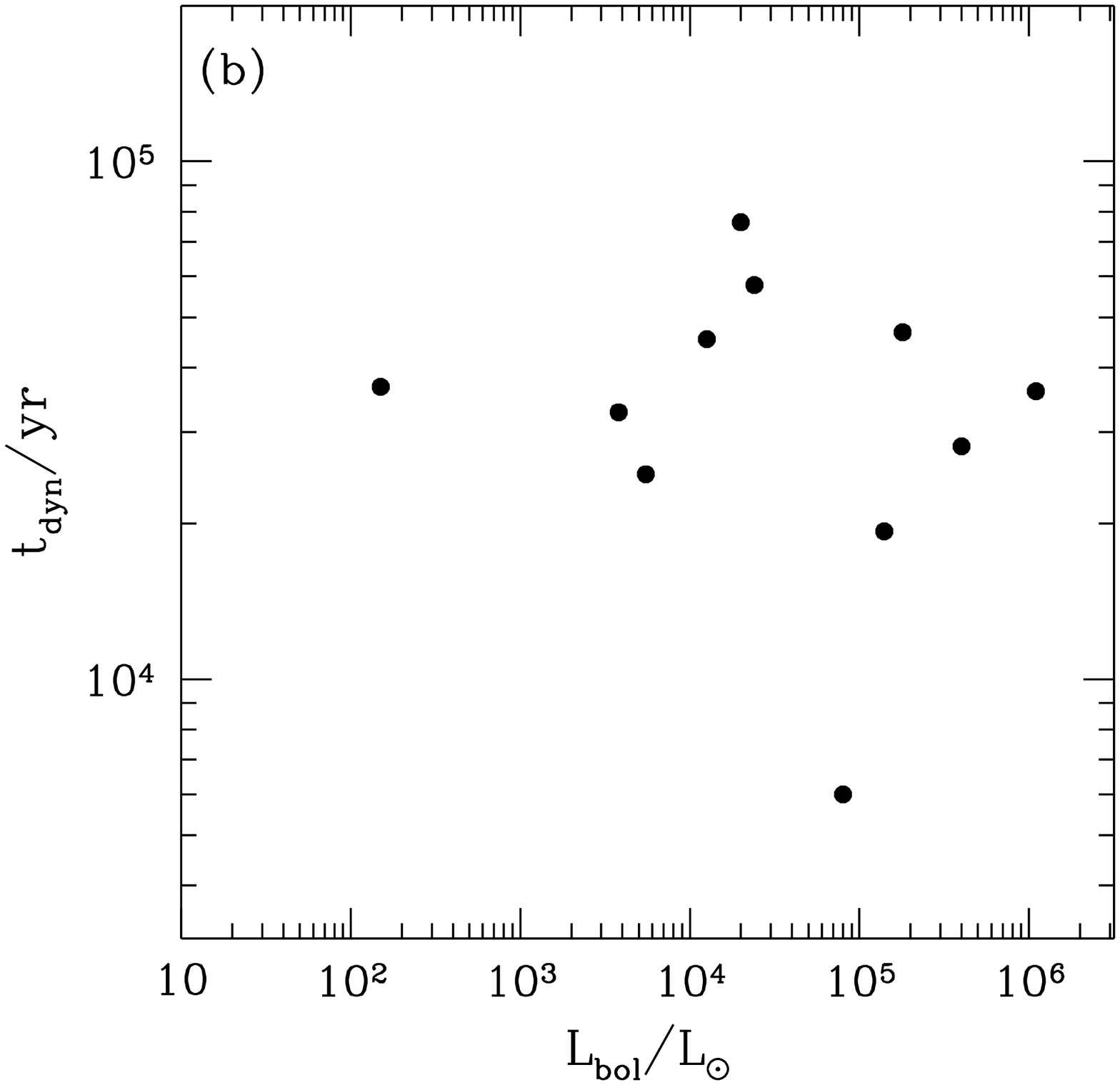}
\caption{(a) Flow size and (b) ``Dynamical timescale''versus 
bolometric luminosity}
\label{rlbol}
\end{figure*}

The sizes of the flow lobes were determined from the contour maps by
measuring the distance in arcsec between the adopted centre of the
flow and the 10\% contour in the direction of the flow axis, where the
flow axis was defined as the straight line joining the blue and red
emission peaks.  The values for the red and blue lobes were then
averaged and converted to a physical size in parsec using the
distances to the sources listed in Table \ref{sdsample}. The dynamical
timescale, t$_{\rm dyn}$ was then calculated by dividing the flow size by
the intensity-weighted velocity. It should be noted however that the
dynamical timescale may not be a true ``age'' of the flow (see \eg\
Padman et al.\ 1997\nocite{pad97} for a discussion), particularly for
sources like AFGL\,437 where there is an obstruction or AFGL\,5142
where the lobes are overlapping. The flow sizes and dynamical
timescales are listed in Table \ref{props}. The flows are all of the
order of 1\,pc in size.
\begin{table*}
\centering
\begin{tabular}{ccccccc}
\hline
  Object & Flow Mass &   \Fco &  \Lco &  Flow Size & Flow age & \.M$_{\rm flow}$\\
         & \Msolar   & $\x10^{-3}$\,\Msolar\,\kms\,yr$^{-1}$& \Lsolar & pc & $\x10^4$yr & \Msolar\,yr$^{-1}$\\ 
\hline
AFGL\,5157 & 672 & 131 & 217 & 0.69 & 2.49 & 0.027 \\ 
AFGL\,5142 & 1082 &10.9& 5.47& 0.80 & 3.28 & 0.033 \\ 
AFGL\,437 & 620 & 235 & 750 & 1.36 &5.77 & 0.011 \\ 
S88\,B & 867 & 103 & 222 & 1.12 & 4.68 & 0.019 \\  
GGD\,27& 407 & 14.7 & 15.8& 1.44& 7.63 & 0.0053\\ 
W3\,IRS5 & 2016 & 756 & 2594&1.73& 3.60 & 0.056 \\ 
NGC\,6334B & 2344 & 14.1 & 6.2 & 0.69& 2.82 &0.083 \\ 
NGC\,6334I & 3317 & 76.6 & 71.9& 0.74& 0.60 & 0.55 \\
IRAS\,20188& 2166 & 2.66 & 1.10& 1.67& 4.54 & 0.048 \\ 
IRAS\,19550& 5.77 &0.29 & 0.35& 0.58& 3.67 & 0.00016\\ 
W75\,N & 548 & 35.2 & 46.4 & 1.23&1.93 & 0.028 \\
\hline
\end{tabular}
\caption{Flow Properties}
\label{props}
\end{table*}
Both R$_{\rm flow}$ and t$_{\rm dyn}$ are uncorrelated ($<\!1\sigma$) with
bolometric luminosity (Fig.\ \ref{rlbol}).

\subsubsection{Flow Mass and Energy}
\label{mef}

Table \ref{props} lists the dynamical age, mass, CO momentum flux
(\Fco) and power (\Lco) we have derived for all the objects in the
sample, following the method of Cabrit \& Bertout (1992), with
corrections applied for variations in optical depth with velocity and
position in the flow (see Shepherd \& Churchwell
1996b\nocite{sc96b}). A full discussion of the procedure is given in
Ridge \cite*{thesis}.

\begin{figure*}
\center
\vspace*{7cm}
\includegraphics{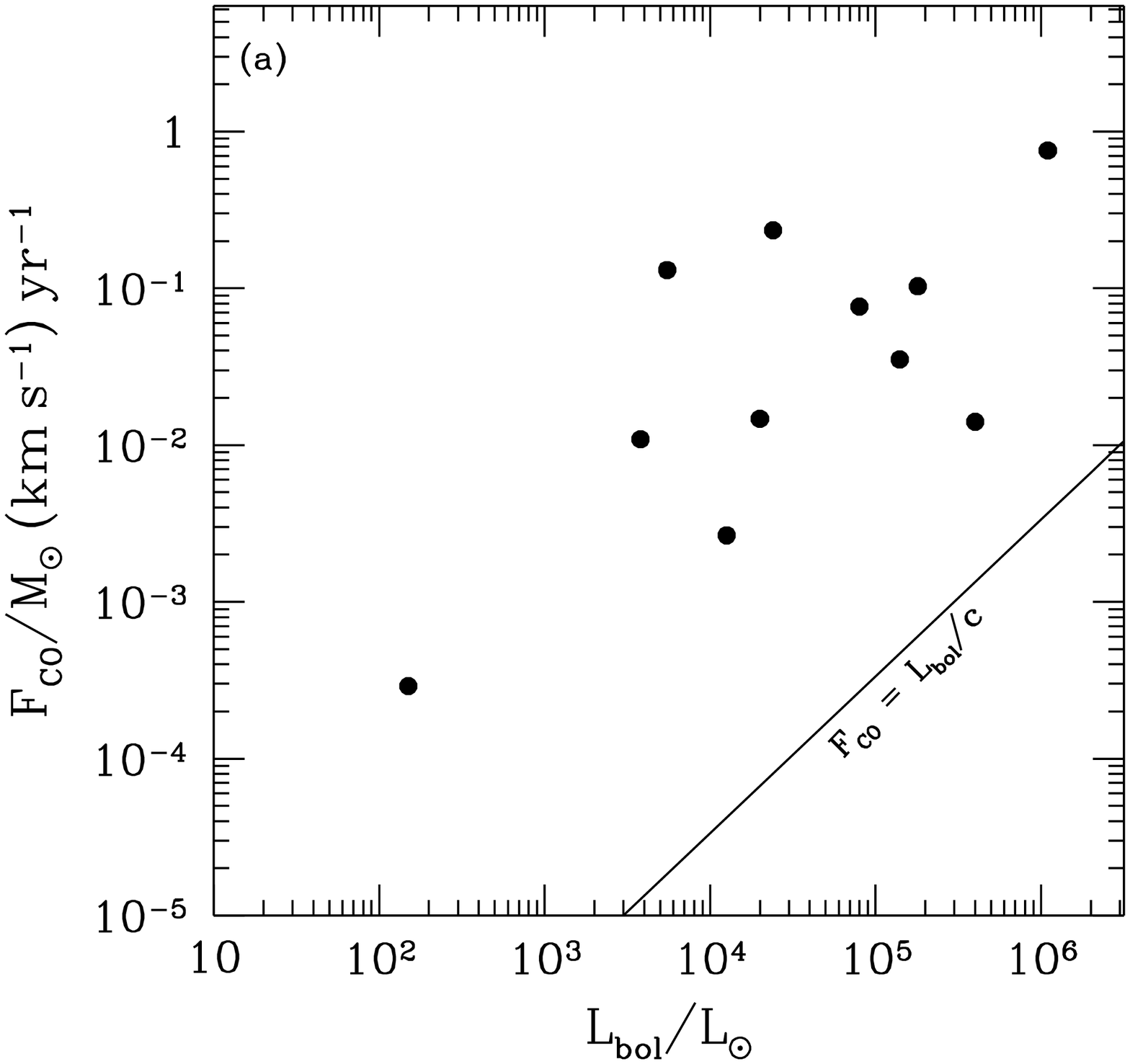}
\includegraphics{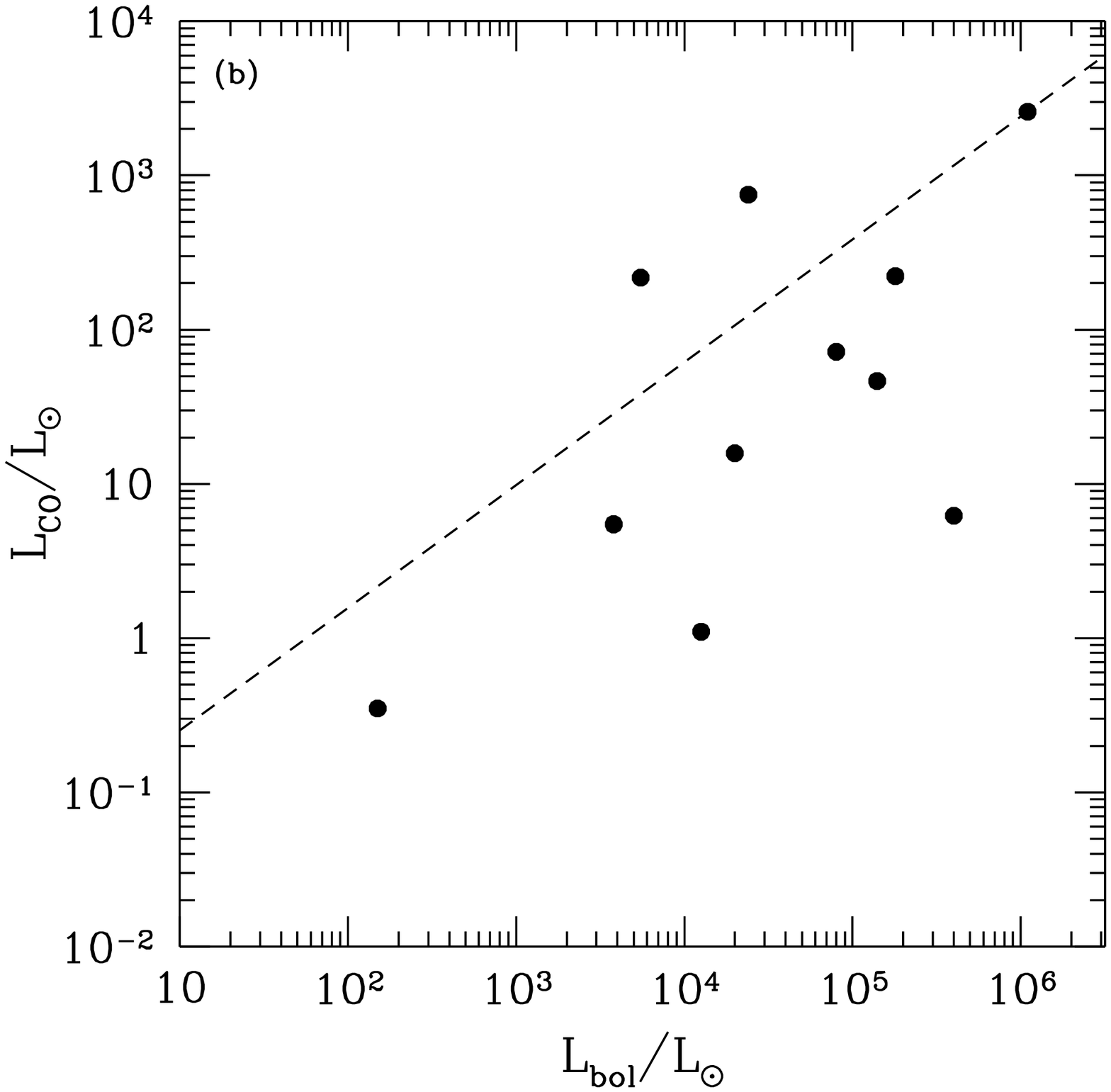}
\caption[Flow force and power vs. bolometric luminosity]
{(a) Flow force and (b) flow power vs. \Lbol\ for the objects in the
single distance sample.  The solid line in (a) indicates the force
available in stellar photons assuming single scattering. The dashed
line in (b) shows the best fit line to the data presented by
Cabrit \& Bertout (1992).}
\label{correlations}
\end{figure*}

Fig. \ref{correlations} shows \Fco\ and \Lco\ plotted against source
bolometric luminosity. Fig. \ref{correlations}a confirms that the force
required to drive the outflows, \Fco, is approximately 100 times
greater than the force available in radiation pressure from stellar
photons (\Lbol/c, assuming single scattering) indicated by the solid
line, and therefore radiation pressure cannot be the sole driving
mechanism for these outflows.
A Spearman rank test gives correlation coefficients of 0.55 between
flow force and \Lbol, and 0.61 between flow power and \Lbol. Neither
of these are equivalent to a significant correlation (2.0$\sigma$ and
2.3$\sigma$ respectively).  There are two particularly noticeable
outliers which fall well below the dashed line indicating the
CB \nocite{cb92} relation in Fig.\ \ref{correlations}.  
The \xco\ emission from the object NGC\,6334B is complex, and may be
optically thick even in the wings so the values of \Lco\ and \Fco\
will be lower limits. Additionally, this flow extended outside the
mapped region due to the poor information available in the literature
about the size and exact location of this flow, causing the mass (and
therefore dynamics) to be underestimated. These factors could easily
increase the measured values of the flow dynamics by up to a factor of
10, bringing it into line with the scatter of the other objects.
There is no similar explanation for the weakness of the flow from
IRAS\,20188, and this source appears to be driving a genuinely weak
outflow.
Additionally, there are two sources which appear to be driving
particularly powerful outflows, AFGL\,437 and AFGL\,5157. However,
both these flows are within a factor of $\sim$5 of CB's
relation, well within the uncertainties expected in deriving dynamical
properties.

Therefore, within the uncertainties involved in calculating dynamical
properties from outflows, only the object IRAS\,20188 is not
consistent with the previously presented relations between outflow
force and \Lbol\ and outflow power and \Lbol.

\subsubsection{Mass Outflow Rate}

\begin{figure}
\vspace*{8cm} 
\includegraphics{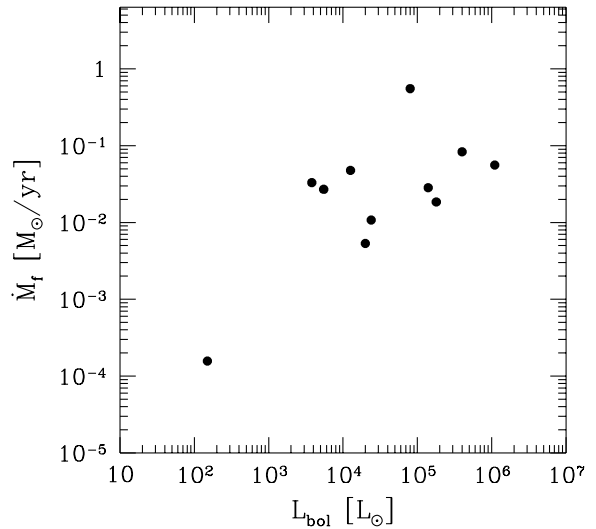} 
\caption{The mass outflow rate \.M$_f$ versus the
bolometric luminosity \Lbol.}  
\label{mdot} 
\end{figure}

Fig. \ref{mdot} shows the mass outflow rate, $\dot M_{\rm f} =
M_{\rm f}/t_{\rm dyn}$ plotted against bolometric luminosity.  The mass
outflow rates calculated for this sample are a factor of $\sim$10
greater than those derived by Shepherd \& Churchwell (1996b) for
objects with a similar bolometric luminosity. This discrepancy is
probably due to the different techniques used for determining the the
velocity at which the mass is dominated by cloud core emission
(low-velocity cut-off in the integration).  Masson \& Chernin
\cite*{mc94} and Ridge \cite*{thesis} have shown that this can
significantly influence the measured mass in a flow, and is the
largest source of error in mass-determination.  Therefore the
low-velocity cut-off has to be selected in a consistent manner when
comparing samples of objects (see Ridge 2000\nocite{thesis} for a
full discussion of the method adopted here).

There is no significant correlation between mass outflow rate and
bolometric luminosity -- a Spearman rank test gives a correlation
coefficient of 0.49, equivalent to a 1.7$\sigma$ correlation. This
appears to be in contradiction with the results of Shepherd \&
Churchwell (1996b). However the relation they present shows a large
scatter in \.M$_{\rm f}$ among the high luminosity objects, and would
probably not produce a significant correlation if the objects in their
sample with \Lbol$<$ 100\Lsolar\ were disregarded.

\subsubsection{Mass-Velocity Relation}

\begin{figure}
\vspace*{6.5cm} 
\includegraphics{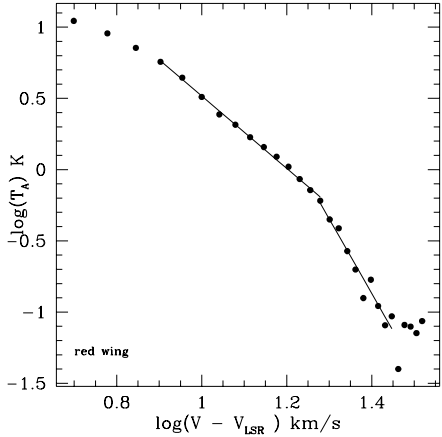} 
\caption{Spatially averaged intensity vs. velocity
for red lobe of W75\,N, showing a power law
relation between intensity and velocity, breaking at $\sim$
18\kms.
}
\label{w75slope} 
\end{figure}

\begin{figure}
\vspace*{6.5cm} 
\includegraphics{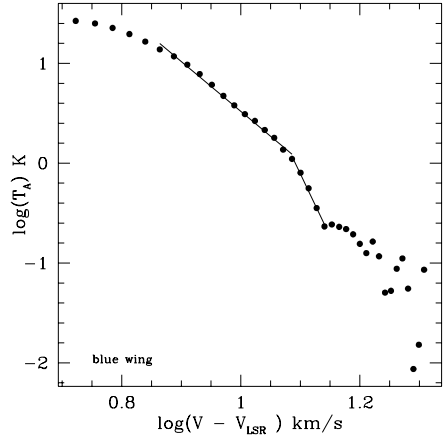} 
\caption{Spatially averaged intensity vs. velocity
for blue lobe of W3\,IRS5, showing a power law
relation between intensity and velocity, at lower
velocities, but a bump at $\sim$16\kms.
}
\label{w3slope} 
\end{figure}

A number of authors have observed a power-law relation between
intensity and velocity in the high-velocity wings of CO (\eg\ Davis et
al.\ 1998, Lada \& Fich 1996).  In most of the objects in this sample,
at least part of the line profile is well described by a power-law
$I\propto V^{-\gamma}$. Least squares fits to the spatially averaged
spectra are summarised in Table \ref{mv}. Where a single power law did
not describe the emission over all velocities, two values are given in
Table \ref{mv}, the first describing the behaviour at low velocities
($\ltsim$10\,\kms) and the second the higher-velocity emission
($\gtsim$10\,\kms).  The slope for the red wing of the object W75\,N
is shown in Fig.\ \ref{w75slope} as an example.
\begin{table}
\centering
\begin{tabular}{ccc}
\hline
Object     & $\gamma_{\rm blue}$  & $\gamma_{\rm red}$ \\
\hline
AFGL\,437  & -5.13$\pm$0.20 & -2.86$\pm$0.03\\ 
AFGL\,5142 & -2.40$\pm$0.07 & -2.19$\pm$0.04\\
AFGL\,5157 & -1.30$\pm$0.01 & -1.59$\pm$0.03\\
           & -2.24$\pm$0.03 & -3.79$\pm$0.14\\
S88\,B     & -9.06$\pm$0.06 & -6.66$\pm$0.09\\
GGD\,27    & -2.60$\pm$0.04 & -3.57$\pm$0.02\\
           &                & -1.55$\pm$0.12\\
W75\,N     & --             & -2.54$\pm$0.02\\
           & --             & -5.23$\pm$0.06\\
W3\,IRS5   & -4.99$\pm$0.02 & -2.90$\pm$0.04\\
           & -12.52$\pm$0.02& -4.15$\pm$0.01\\
IRAS\,19550& -1.91$\pm$0.04 & -1.52$\pm$0.03\\
IRAS\,20188& -1.21$\pm$0.01 & -3.74$\pm$0.06\\
           & -5.07$\pm$0.08 & \\
NGC\,6334I & -0.98$\pm$0.03 & -0.99$\pm$0.06\\
           &                & -9.23$\pm$0.4\\
\hline
\end{tabular}
\caption{The value of $\gamma$, the power-law exponent of the 
mass-velocity relation for the objects in the single-distance sample.
Left column gives the value for the blue wing ($\gamma_{\rm {blue}}$)
and right column the value for the red wing ($\gamma_{\rm {red}}$).}
\label{mv}
\end{table}
There is large spread in the values of both the
low- and high-velocity $\gamma$s measured for the flows.  The
mass-spectrum slope is steeper for the high-velocity emission in all
cases except the red wing emission of GGD\,27.  The blue wing of
W3\,IRS5 shows particularly unusual behaviour, with the mass-velocity
relation steepening to a slope $\sim -12$ at a velocity of
$\sim$10\,\kms, then showing a small bump between velocities of 14 and
16\,\kms\ (see Fig.\ \ref{w3slope}).

\begin{figure}
  \vspace*{15cm}
\includegraphics{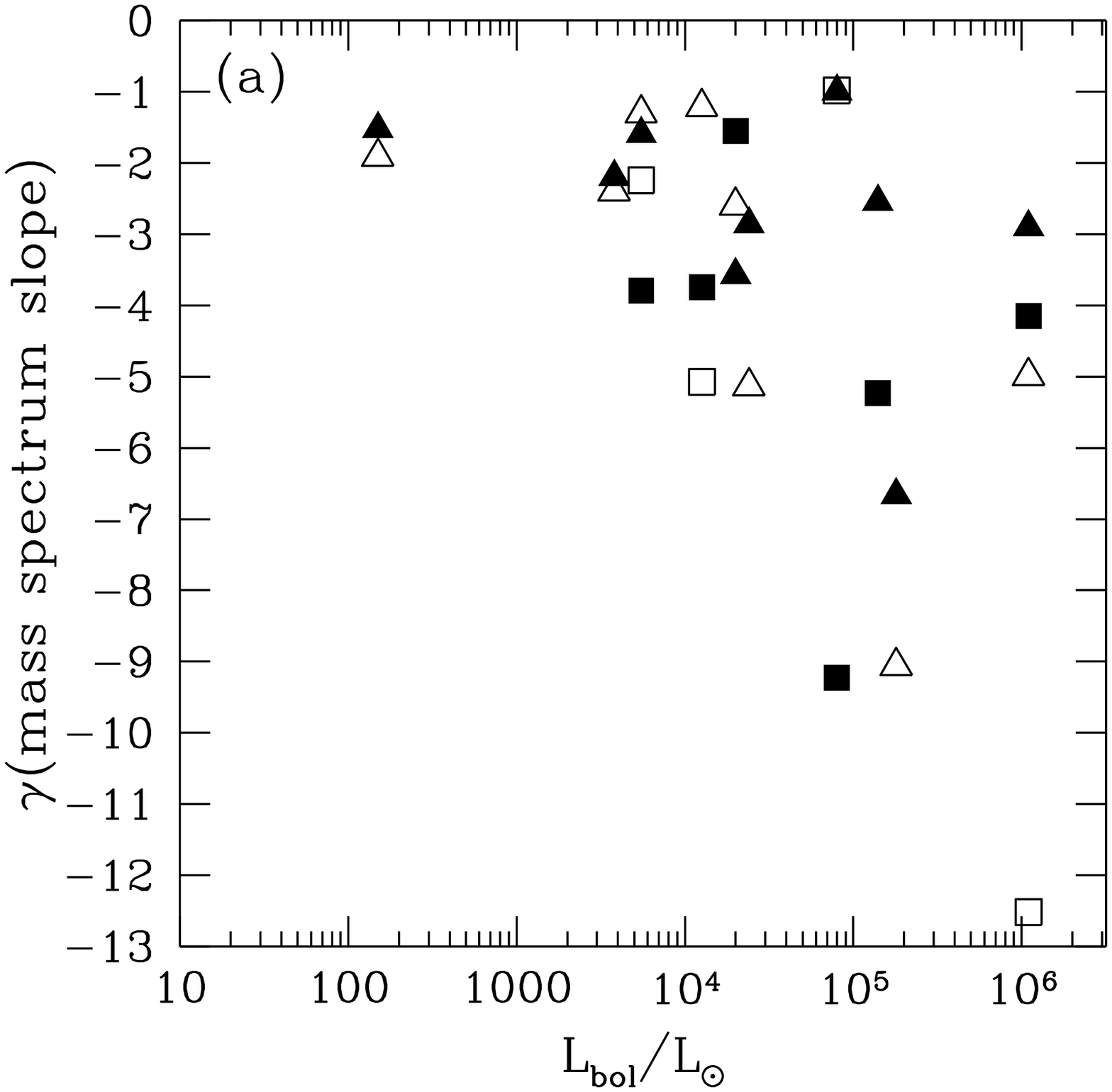}
\includegraphics{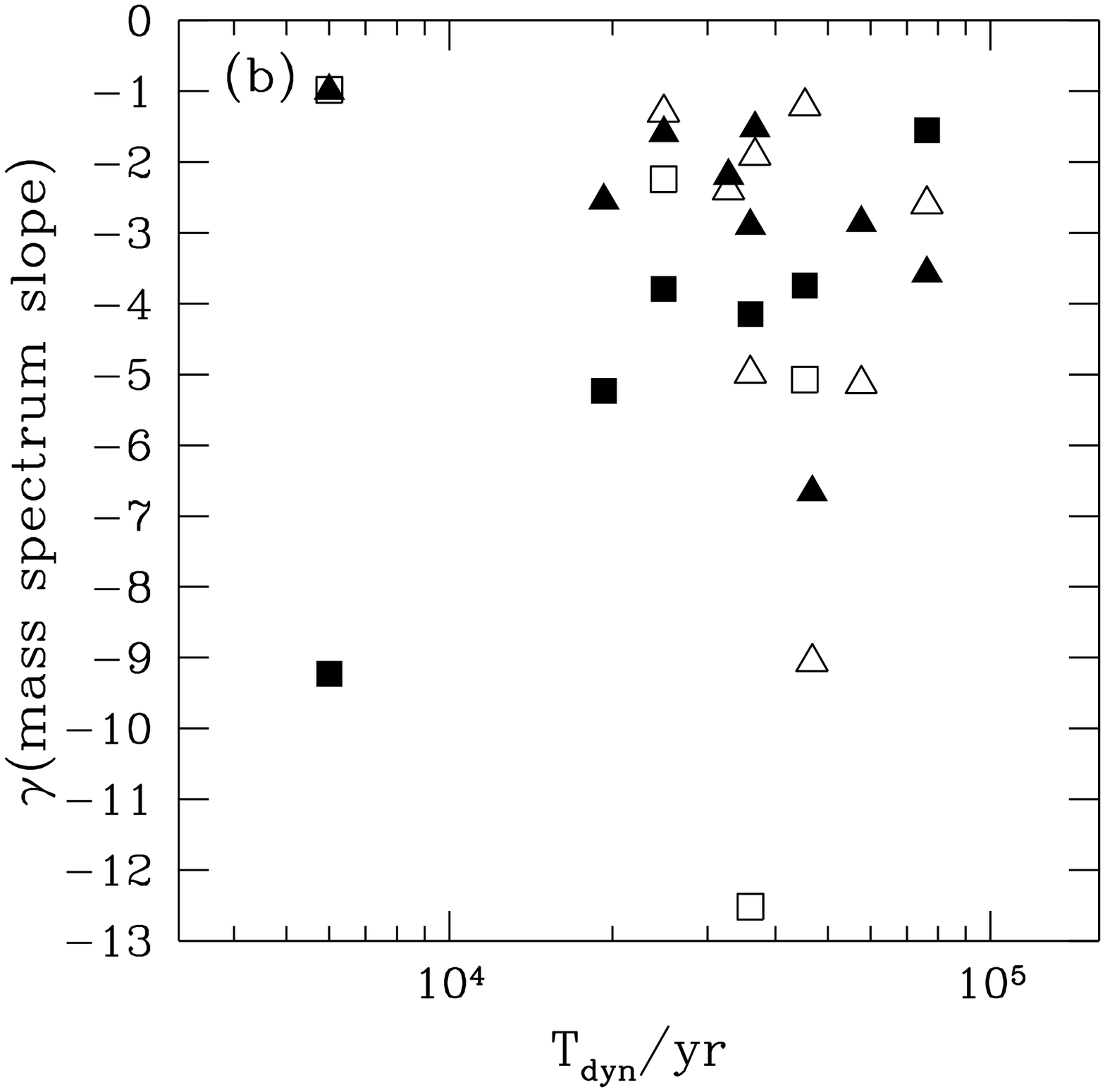}
\includegraphics{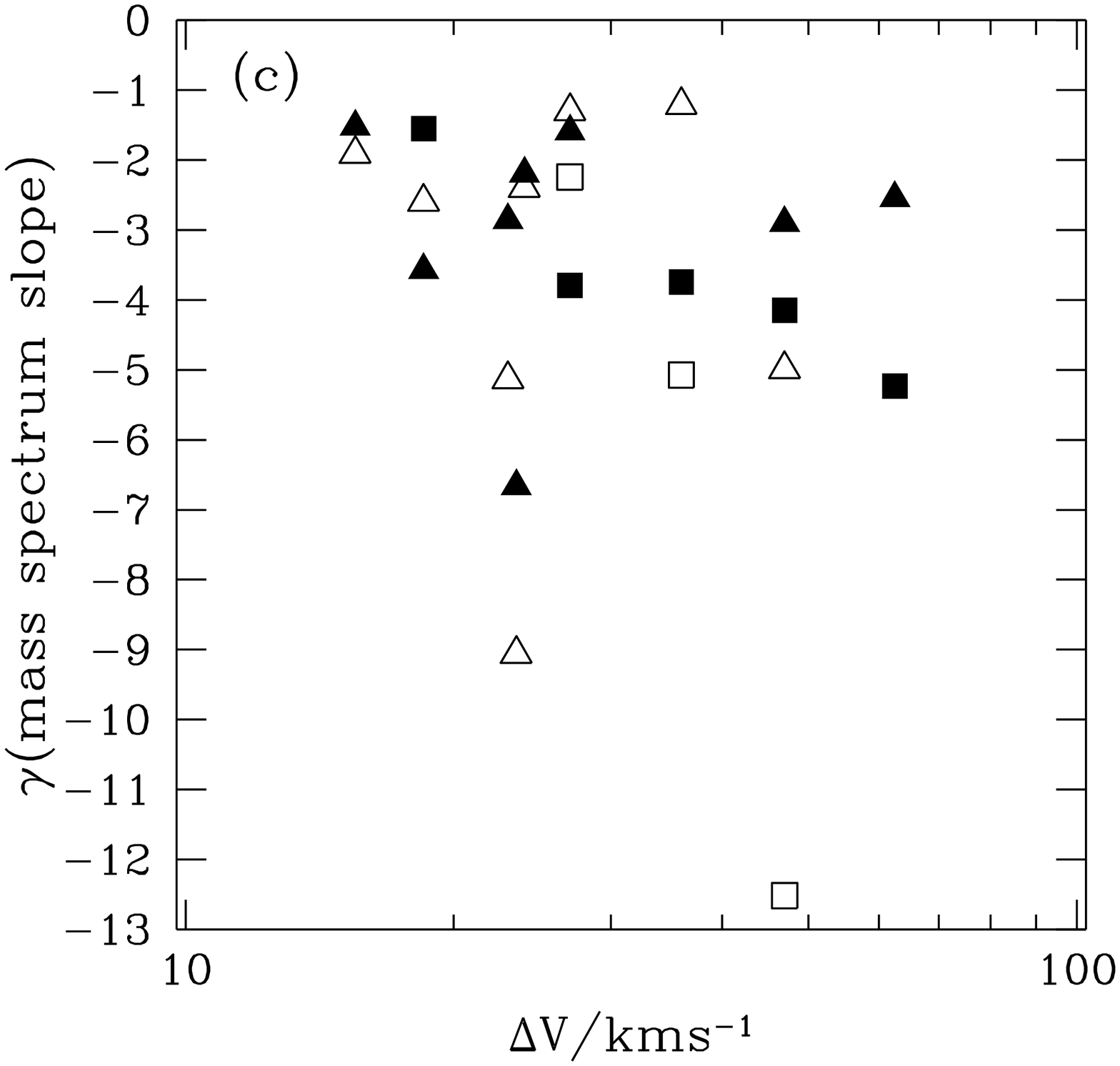}
\caption[Mass spectrum slope vs. flow properties] 
{The slope $\gamma$ of the mass
  spectrum $M(v)$ plotted as a function of (a)~source bolometric
  luminosity \Lbol, (b) dynamical age t$_{\rm dyn}$ and (c) wing extent
  $\Delta V$.  Triangles represent $\gamma$ for gas with low velocity
  ($\ltsim$10\,\kms) relative to the source, and squares are for gas
  moving at more than 10\,\kms. Filled symbols show $\gamma$s measured
  for red shifted emission, and open symbols show $\gamma$s for
  blue-shifted emission.}
\label{gam}
\end{figure}

The low-velocity mass-spectrum slopes observed for the majority of the
flows are significantly steeper than the often-quoted value of $\sim$
1.7--1.8 for low-mass YSO outflows \cite{lf96}. This supports the
conclusions of Davis et al.\ (1998) who also found that outflows from
high-mass YSOs have a steeper mass-velocity relation than is observed
in low-mass flows. They interpreted this as evidence for the more
evolved nature of the flows they observed. We do not find a
correlation between mass-spectrum slope and bolometric luminosity
however (see Fig.\ \ref{gam}a).

The slopes derived here cannot be considered a true mass-velocity
relation, as in most cases, the emission is optically thick, and
therefore mass is not directly proportional to intensity.  The optical
depth decreases with velocity (Ridge 2000, Shepherd \& Churchwell
1996b) and therefore the correction is more extreme for the lower
velocity emission.  This means that the true slope ($\gamma$) of the
mass-velocity relation for these objects would be even steeper.  Table
\ref{mv} also shows that in all but one case (GGD\,27), the value of 
$\gamma$ for the red wing is smaller than $\gamma$ for the blue wing.
It is not clear what the cause of this effect is. Optical depth
effects would cause red-shifted self-absorption in infalling material
(e.g. Myers et al.\ 1996\nocite{myers96}).

Fig.\ \ref{gam} shows the slopes plotted against bolometric
luminosity, dynamical timescale t$_{\rm dyn}$ (flow age) and wing
extent. We do not find the clear separation between $\gamma$s measured
for the high velocity ($\gtsim$ 10\,\kms) and low velocity emission
that was presented by Richer et al.\ \cite*{richer00} for a sample of
twenty-one outflows spanning a luminosity range from
$\sim$0.3--10$^6$\Lsolar\ compiled from the literature.  There is some
suggestion that there is an increase in the magnitude of $\gamma$ with
luminosity, but there are too few sources in the low-luminosity range
to make this conclusive. The spread of $\gamma$s does seem to increase
with bolometric luminosity, similar to Fig.\ 4 of Richer et al.\
\cite*{richer00}.  They also proposed that there exists a
weak relationship between outflow dynamical age and mass-spectrum
slope in high mass YSO outflows, with the slope becoming steeper with
age. This is not evident in Fig.\ \ref{gam}b, although we only have a
relatively narrow range of flow ages so cannot rule this possibility
out completely.  Fig.\ \ref{gam}c shows $\gamma$ plotted against
$\Delta V$, to test whether the value of $\gamma$ is affected by the
detectable width of the line wings. No correlation is present between
either the blue- or red-shifted $\gamma$s and $\Delta V$ ($\ll
1\sigma$ in all four cases).

\section{Discussion}

\subsection{The Sample}

The objects in this single-distance sample were selected from the
catalogue of Wu et al.\ (1996). This catalogue was compiled from the
literature and includes YSOs identified from their IRAS spectral
energy distribution (SED), radio surveys for H\,{\sc ii} regions (\eg
Wood \& Churchwell \nocite{wc89})and radio-continuum emission, and
other outflow objects discovered either serendipitously or as part of
targeted observations towards known YSO candidates. The catalogue
cannot therefore be considered complete, and is not volume-limited.
Wu et al.\ do not discuss the possible selection effects in their
catalogue. Bearing this in mind, the sample may not be representative
of the true YSO population. 

In particular, there may be biases in our sample associated with YSOs
identified from their IRAS SED and/or existence of radio
continuum. This is likely to bias the selection to older YSOs (Class I
or later), as many Class 0 objects are still invisible even at
far-infrared wavelengths. Outflows found as part of targeted
observations of known YSOs may also introduce biases, as it is often
the interesting, and therefore probably unusual (strongest), objects
that are selected for detailed multi-wavelength study.

However, by selecting objects with a range of luminosities with the
same distance, we believe we have achieved a good approximation to a
representative sample of the properties of YSO outflows.

\subsection{Correlation between \Fco\ and \Lbol}

We find a marginal (2.3$\sigma$) correlation between outflow power and
source luminosity (section \ref{mef}). A least squares fit shows a
relation of \Fco $\propto$ \Lbol$^{0.62\pm0.21}$, which is consistent
with the slope of 0.69$\pm$0.05 for the high-luminosity objects in the
sample of CB \nocite{cb92}. Our data are offset to lower \Lco\
compared to Cabrit \& Bertout's data, but again are not inconsistent
within the uncertainties. Cabrit \& Bertout selected well-collimated
flows for their study, and it is therefore likely to be dominated by
outflows from Class 0 sources \cite{bontemps96}.  However, YSO classes
are only defined for low-mass objects and just a handful of high-mass
Class 0 candidates have been identified to date (\eg Molinari et al.\
1998\nocite{molinari98}). Models of high-mass star-formation (\eg
Stahler, Palla \& Ho 2000 \nocite{stahler00}) predict there may in
fact be no such thing as a high-mass protostar, as the stellar
birthline intersects the main-sequence for stars of
$\gtsim$10\Msolar. It is therefore likely that the majority of the
sources in our sample are equivalent to Class I objects or
later. Bontemps et al.\ \cite*{bontemps96} find a steeper slope
($\sim$1) for the Class I sources in their low--intermediate mass
(\Lbol $<$100) sample (consistent with the slope for just the
lower-mass (\Lbol $\ltsim$ 100) objects in CB's \nocite{cb92}sample),
giving an almost linear relationship between \Fco\ and \Lbol, and
indicating that all Class I sources have approximately the same
outflow efficiency,
\Fco/F$_{\rm rad}$, of $\sim$100, where F$_{\rm rad}$ is the radiative
momentum flux, equal to \Lbol/c. Bontemps et al.\ also find that Class 0
sources have an outflow efficiency of $\sim$1000,
\ie a factor of $\sim$ 10 more efficient than Class Is.

\begin{figure}
\vspace*{7cm}
\includegraphics{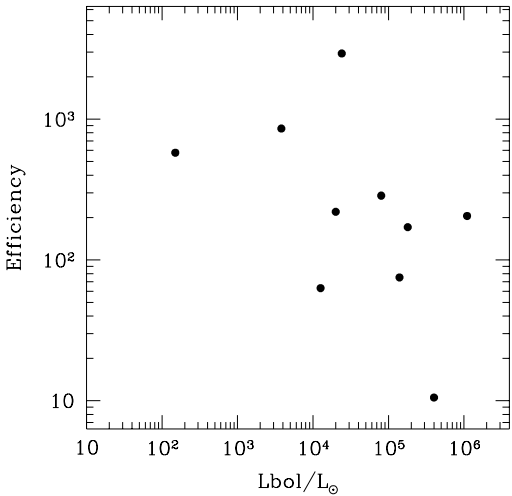}
\caption[Outflow efficiency versus \Lbol]
{Outflow efficiency versus \Lbol\ for the objects in the single
distance sample}
\label{eff}
\end{figure}

The outflows in our sample show a range of outflow efficiencies (Fig.\
\ref{eff}) from $\sim$10 to $\sim$ few times $10^3$ (\ie 10 to 1000 
times the force due to radiation pressure assuming single scattering),
too large for radiation pressure to be important in driving these
flows, even if multiple scattering is present.
If outflow efficiency is an indicator of YSO class (and therefore age)
as Bontemps et al.\ suggest based on their observational results, then
the single-distance sample must contain a mixture of Class 0 and Class
I sources. There is a marginal (2.1$\sigma$) anti-correlation between
outflow efficiency and bolometric luminosity in our sample. This may be
an evolutionary effect, as massive YSOs are thought to evolve more
rapidly, and it is therefore likely that on average we are seeing them
at a later stage in their pre-main-sequence evolution.

The outflow efficiencies agree well with other previous works
covering several decades in luminosity (\eg
Rodr{\'\i}guez et al.\ 1982\nocite{rod82}, Bally \& Lada 1983\nocite{bl83}, 
Lada 1985\nocite{lada85}, Levreault 1988\nocite{lev88}) and with 
CB \nocite{cb92} who found an efficiency of $\sim$250 for their entire 
sample, while for the lower luminosity sources alone the efficiency 
was $\sim$1000.

\subsection{Collimation}

The flows in the single distance sample are much less well collimated
than the molecular outflows from their low-mass counterparts
(collimation factors, ${\rm R_{coll}=R_{max}/R_{min}}$, of $\sim$1--2 in the
high-mass sample as opposed to $\gtsim$6 in low-mass objects
\cite{lada85}). There could be several reasons which could explain the
relatively poor collimation of these flows.  One possibility is that
in some cases of high-mass outflows we are not seeing the outflow from
a single star, but the superposition of several highly-collimated
outflows from a cluster of young stars. High-mass star forming regions
are also generally further away than the well-studied low-mass star
forming regions, and therefore the hypothesis of multiple outflows
requires high-resolution observations in both the mid-infrared (to
resolve the stellar cluster) and in \co\ (to resolve the outflows) in
order to be tested thoroughly. The \co\ observations would require
mm-interferometry well beyond the capabilities of current facilities.
Alternatively, the molecular outflow may be tracing the density
structure close to the YSO. For instance, Mellema \& Frank
\cite*{mf97} have shown that aspherical bubbles can form from the
interaction of a central YSO wind with a toroidal circum-protostellar
density distribution. Observations of cloud core morphologies
\cite{thesis} do not seem to indicate a toroidal morphology however.

\subsection{Mass-Velocity Relation}

We find an extremely steep mass-velocity relation in a number of the
flows in the sample, indicating a high mass-fraction at lower
velocities. This has previously been interpreted as evidence for the
more evolved nature of high-mass outflows, as the majority of material
has had time to slow down after its initial acceleration due to either
prompt or turbulent entrainment (\eg Davis et al.\ 1998
\nocite{davis98b}). However it is also possible that high-mass
outflows are accelerated by a different mechanism, with the majority
of the momentum transferred to large amounts of slow-moving material
in the flow.

Six of the eleven objects in the sample show distinct breaks in the
power-law between low- and high- velocity gas, as has been observed in
some low-mass flows (\eg Lada \& Fich 1996\nocite{lf96}).  Richer et al.\
\cite*{richer00} found that the slopes ($\gamma$s) for the
low-velocity gas are similar in sources of all luminosities
(0$<\gamma<$2.5), and interpreted this as further evidence that a
common acceleration mechanism operates over nearly six decades in
\Lbol. Richer et al.\ also found a clear separation between $\gamma$s
in high- and low-velocity gas, interpreting this as evidence for two
distinct outflow velocity components, corresponding to a recently
accelerated component and a slower, coasting component. We found a
large range of $\gamma$s for both the low- and high-velocity gas, and
therefore my data does not support these conclusions.

Models of jet-driven molecular outflows (\eg Downes \& Ray 1999
\nocite{downes99}, Smith et al.\ 1997) \nocite{smith97} have 
tried to reproduce the power-law of $\sim$1.7 observed by \eg Lada \&
Fich \cite*{lf96} in low-mass flows with some success, and the
hydrodynamic simulation by Smith et al.\ \cite*{smith97} also
predicted the change in slope at high velocities, due to a jet-bow
shear layer consisting of molecules which survived the jet terminal
shock.  Smith et al.\ also predict that $\gamma$ should steepen over
time, possibly due to the collection of a reservoir of low-velocity
gas. Unlike Richer et al. we do not find any correlation between
outflow age (dynamical timescale) and the magnitude of $\gamma$, but
the data contain only a narrow range of flow ages, so we cannot rule
this possibility out completely.

Alternatively, Downes \& Ray \cite*{downes99} state that in their model, the
molecular fraction in the jet has a significant influence on the value
of $\gamma$ their model predicts, with $\gamma$ increasing with
decreasing molecular abundance in the jet. If this is the case, then
the high values of $\gamma$ found for these high-mass outflows would
imply a small molecular fraction in the entraining jet, and the large
range of $\gamma$s ($\sim 1$ to $\sim 12$) would suggest that a
large range of molecular abundances is possible in protostellar jets.
However, their simulation of an atomic jet, which should give an upper
limit to the value of $\gamma$ their model can produce, gives $\gamma$
of 3.75, much less than the observed values even for the low-velocity
material, and it is therefore unclear whether the molecular abundance
of the jets has any role to play in the slope of the mass-velocity
relation.

\subsection{The Hubble Law}

The existence of a Hubble-like relation ($v\propto r$) for molecular
outflows (\eg Lada \& Fich 1996\nocite{lf96}) has been challenged by
Padman et al.\ \cite*{pad97}, who state that such a law may apply {\em
locally} at particular bow shocks, and this could be interpreted as a
more global phenomenon in small or incompletely-mapped sources. This
conclusion is supported by the numerical simulations presented by \eg
Downes \& Ray \cite*{downes99} which show that the Hubble law is
almost certainly a local effect in the vicinity of a bow shock.  If a
Hubble law was more global, it would be difficult to reconcile with
jet-driven models which accelerate the molecular material {\em in
situ} rather than it originating at the source. The observations
presented here show no evidence for a Hubble law in any of the
high-mass objects studied (see also the channel maps presented in
Ridge 2000).

Models such as the simulations of Smith et al.\ \cite*{smith97} reproduce
Hubble-like position-velocity diagrams well. However, the lack of
observational evidence for a global Hubble-like flow puts this
phenomenon in doubt.

\section{Conclusions}
Using a sample of known outflow objects at a single distance, we have
investigated the physical and dynamical properties of intermediate and
high-mass YSOs. Our main results are:
\begin{itemize}
\item We find a similar relation between \Fco\ and \Lbol\ as previous
studies (CB \nocite{cb92}, Shepherd \& Churchwell 1996b\nocite{sc96b}) but with a
large scatter.

\item We find a much steeper mass-velocity relation than has been observed
in low-mass outflows. There is also a large range of values of $\gamma$,
from $\sim -1$ to $\sim -12$.

\item None of the objects studied show a ``Hubble-law'' like relation.
\end{itemize}

We have found that several of the often-quoted observational results
for low-mass YSO outflows are not applicable to high-mass stars. To
attempt to understand fully high mass star formation and their
outflows, more extensive statistical studies need to be undertaken, as
we have shown that the extrapolation of outflow properties from
low-mass stars do not necessarily agree with observations of high mass
stars.

\begin{acknowledgements}
We would like to thank John Porter for his comments on earlier
versions of this manuscript, Chris Collins for useful discussions and
the referee Rafael Bachiller for useful suggestions. NAR acknowledges
the support of a PPARC studentship.  The James Clerk Maxwell Telescope
is operated by the Joint Astronomy Centre on behalf of the Particle
Physics and Astronomy Research Council of the United Kingdom, the
Netherlands Organisation for Scientific Research and the National
Research Council of Canada. The National Radio Astronomy Observatory
is a facility of the National Science Foundation, operated under
cooperative agreement by Associated Universities, Inc.
\end{acknowledgements}

\bibliographystyle{aabib99}
\bibliography{refs}

\end{document}